\documentclass[12pt,a4paper]{article}
\usepackage[T1]{fontenc}
\usepackage{fancyhdr}
\usepackage{fullpage}
\usepackage{amsmath}
\usepackage{accents}
\usepackage{amssymb}
\usepackage{graphicx}
\usepackage[style=authoryear,backend=bibtex,maxcitenames=2,maxbibnames=99]{biblatex}
\bibliography{biblio}
\DeclareNameAlias{sortname}{last-first}
\makeatletter
\AtEveryBibitem{%
  \global\undef\bbx@lasthash%
  \clearfield{}}
\makeatother

\usepackage{epic,eepic}
\usepackage{graphicx}
\usepackage{authblk}
\usepackage{fullpage}
\usepackage[active]{srcltx}
\usepackage{enumitem}
\usepackage{booktabs}
\usepackage[left=1.5cm,right=1.5cm]{geometry}
\usepackage{multirow}

\usepackage{framed,color}

\usepackage{float}%
\floatstyle{plaintop}%
\restylefloat{table}

\usepackage{hyperref}

\date{}
\title{\textbf{Latent Position Network Models}}
\author[1,2]{Hardeep Kaur}
\author[1]{Riccardo Rastelli}
\author[1,2]{Nial Friel}
\author[3]{Adrian E. Raftery}
\affil[1]{\footnotesize School of Mathematics and Statistics, University College Dublin, Ireland;}
\affil[2]{\footnotesize Insight: Centre for Data Analytics, University College Dublin, Ireland;}
\affil[3]{\footnotesize Department of Statistics, University of Washington, Seattle, USA.}

\begin{document}
	
	\maketitle
\begin{abstract}
\noindent
In this chapter, we present a review of latent position models for networks. We review the recent literature in this area and illustrate the basic aspects and properties of this modeling framework. 
Through several illustrative examples we highlight how the latent position model is able to capture important features of observed networks. We emphasize how the canonical design of this model has made it popular thanks to its ability to provide interpretable visualizations of complex network interactions. We outline the main extensions that have been introduced to this model, illustrating its flexibility and applicability. 
\\

\end{abstract}

\section{Introduction}
Statistical network analysis has recently emerged as a prominent area of research, with applications in many fields including social sciences, biology, finance and physics.
Social networks are used to study actors and the pairwise interactions between them. 
The formulation of statistical models for such network data plays an important role in describing the network's global topology and in providing interpretable representations of the data. 
Some of the pioneering work in network analysis dates back to \textcite{erdos59a}, who introduced the famous Erd\H{o}s-R\'{e}nyi model with two connected variants for the generation of random graphs or the evolution of a random network. 
An initial version of what are called today latent position models was developed soon afterwards by \textcite{gilbert61}, who introduced the so-called spatially embedded random networks. 
The key aspect of these models is that they define a generative framework for the edges of a graph based on the positions of the nodes in a Euclidean space. 

\textcite{Hoff20021090} adapted similar concepts and ideas to the analysis of social networks, by introducing a new framework called the Latent Position Model (LPM) or Latent Space Model (LSM).
This work renewed interest in spatially embedded models as a tool for modelling the complex networks arising in many applied fields. 
As a result, the literature on the topic has increased rapidly, and the LPM has become a widely used statistical model for network analysis. 
The model has been used in the analysis of 
corporate governance \parencite{Friel20166629}, 
financial risk \parencite{Tafakori2021}, 
trophic food webs \parencite{Chiu2014139, Chiua201115881}, 
protein sequence data \parencite{Ding2019}, 
interbank networks \parencite{Linardi2019},
trade networks \parencite{Ward201395}, 
social influence \parencite{Sweet2020251, mcfowland2021estimating}, 
music contests \parencite{angelo2019900}, 
migration flows \parencite{xiao2022complex},
anomaly detection \parencite{lee2021anomaly}, 
political networks \parencite{ng2021modeling}, 
conflict networks \parencite{westveld2011mixed}, 
flows of controlled substances \parencite{berlusconi2017determinants},
among others.
It has also been used in a variety of other research areas, including
neuroscience \parencite{Durante20171547, Wilson2020384}, 
item response theory \parencite{Jin2019236},
mediation analysis \parencite{liu2021social}, 
education research \parencite{Sweet2013295}, 
and epidemiology \parencite{chu2021dynamic}.

The literature on LPMs has also been reviewed in previous articles on network analysis, including \textcite{Rastelli2016407, Salter-Townshend2012243, matias2014modeling, Raftery20171531, Smith2019428, Kim2018105, Sosa2021171}.
This paper provides a review of LPMs and of the recent literature that originated from the seminal paper of \textcite{Hoff20021090}.
We include an explanation of the modelling framework in Section~\ref{sec:lpm_framework}, and an application to an original dataset in Section~\ref{sec:example}. This is followed by a detailed literature review in Section~\ref{sec:literature_review}.

\section{Notation}
Throughout this paper, we consider data on the relations between a set of nodes $N \in \{1,\dots,n \}$. 
The relations are described by the set of edges $\mathcal{E}$. 
We denote by $Y$ = $(y_{ij})_{1\leq i < j \leq n}$ the $n\times n$ adjacency matrix of the observed undirected network, where $y_{ij}$ is the edge value between the nodes $i$ and $j$. These edge values are usually binary, discrete or continuous. 
A collection of dyad-specific covariates may be available.
We denote these by $X = (x_{i,j,k})$, where $(i,j)$ is an ordered pair of nodes and $k = 1, \dots, p$ indexes the covariates.
We denote by $\boldsymbol{\theta}$ the collection of parameters that do not refer to edges or nodes; we call these the global parameters. We denote by $Z$ the unobserved latent variables, i.e. the latent positions associated with the nodes. Here $z_{i}$ represents the unknown position of the corresponding node in a $d$-dimensional latent space.

\section{Latent Position Models}\label{sec:lpm_framework}
\subsection{The original distance model and projection model}
Latent position models (also called latent space models) were introduced by \textcite{Hoff20021090}, for undirected binary networks. 
This introduced a new family of latent variable models for network data, with the goal of providing a visualization of a relational dataset through a latent geometric social space.
Crucially, the authors introduced an inferential framework to estimate this model for networks of moderate size (of around a few hundred nodes).

The social space introduced by the authors consists of a set of $n$ nodes, each having an unknown position $z_i$ in a $d$-dimensional latent space, typically $\mathbb{R}^d$. 
In a Bayesian framework, the positions are assumed to be independent and identically distributed according to a multivariate Gaussian distribution $\mathcal{MND}(\textbf{0}, \Sigma)$.
One fundamental property of this model is that, conditionally on the latent positions $Z$, the relational ties $y_{i,j},\ \forall$  $i,j= 1,\dots,n$ are independent. 
In particular, the probability of the presence of a generic edge $y_{i,j}$ is modeled as some function of the latent positions $z_i$ and $z_j$ of the two nodes involved, as shown in Eq \ref{eq:lpm_hoff_1}. 
\begin{equation}\label{eq:lpm_hoff_1}
\mathbb{P}(Y|Z,X,\boldsymbol{\theta}) = \prod_{i < j} \mathbb{P}(y_{i,j}|z_i,z_j,x_{i,j},\boldsymbol{\theta}),
\end{equation}
where the product is taken over all pairs $(i,j)$, where $i<j$, for $i,j=1,\dots,n$.

Thanks to their geometric framework, these models are naturally able to represent common network features such as reciprocity, transitivity, and homophily. 
In other words, nodes with similar characteristics (i.e. positions) tend to possess higher probability of forming a tie.
The similarity of latent characteristics may be defined in different ways: the two approaches proposed by \textcite{Hoff20021090} are called the distance model and the projection model.

\subsubsection{Distance model}
The latent distance model proposed by \textcite{Hoff20021090} assumes that each node has an unobserved latent position $z_i$ in an Euclidean latent space, typically in the plane $\mathbb{R}^2$. Then, the closer two nodes' latent positions  are to each other, the higher the probability that there is a connection between them.
Conversely, the probability of connection decreases as the distance between the nodes increases. 
The Euclidean distance is most commonly used, but any other distance function may be considered. 
This framework permits easy visualization and interpretation of the latent social space. The formulation of \textcite{Hoff20021090} defines the log odds of a tie between nodes $i$ and $j$ as:
\begin{equation}\label{eq:lpm_hoff_2}
\begin{aligned}
	\eta_{i,j} = {\rm logodds}(y_{i,j}=1|z_i,z_j,x_{i,j},\alpha, \beta) = \alpha + \beta^{\prime} x_{i,j} - |z_i - z_j|, 
\end{aligned}
\end{equation}
where $\alpha$ and $\beta$ are real-valued and $|\cdot|$ indicates the Euclidean norm.

This parameterization is for a logistic regression model with $\boldsymbol{\theta} = (\alpha, \boldsymbol{\beta})$ where $\alpha$ is an intercept term and $\boldsymbol{\beta}$ is a vector of coefficients for covariate effects. The distance model is particularly suited for networks with undirected or directed relations exhibiting strong reciprocity.

\subsubsection{Projection model}
\textcite{Hoff20021090} postulated the projection model as an alternative to the distance model. 
This proposes that two nodes have a higher probability of connecting if their respective latent positions are in the same direction $(z_i'z_j > 0)$, or conversely, they are less likely to connect if they point in opposite directions $(z_i'z_j < 0)$. In other words, two nodes are likely to form a tie if the angle (with respect to the center of the space) between them is small, and less likely to form a tie if the angle between them is obtuse. The projection model is defined as follows:
\begin{equation}\label{eq:lpm_hoff_3}
\begin{aligned}
	\eta_{i,j} = {\rm logodds}(y_{i,j}=1|z_i,z_j,x_{i,j},\alpha, \beta) = \alpha + \beta^{\prime} x_{i,j}- \frac{|z_i^{\prime}. z_j|}{|z_j|} 
\end{aligned}
\end{equation}
The latent effects depend on $\frac{|z_i^{\prime} z_j|}{|z_j|}$ which is the signed magnitude of the projection of $z_i$ in the direction of $z_j$. 
This quantity can be interpreted as the extent of shared characteristics among nodes $i$ and $j$, multiplied by the activity level of $i$.

Similarly to the distance model, the projection model provides a latent view of the social space.
The latent space, in this case, is more easily thought of in terms of polar coordinates, where the position of a node is interpreted as a direction and a magnitude, indicating the social orientation and sociality effect, respectively.
While the distance models naturally yields only symmetric edge probabilities, the projection model can also represent asymmetric edge probabilities, and may thus be more suitable for directed networks.
However, it should also be noted that the distance model typically provides a more clear and intuitive representation of the latent space, and arguably more flexibility in representing some network topologies, such as community structures.

\subsection{Latent distance models with clustering}
An extention of the latent space model was proposed by \textcite{Handcock2007301}, who introduced the Latent Position Cluster Model (LPCM). Their innovation was to model the clustering of highly connected nodes in the network via a latent mixture mode.  This model has been extended by others, including \textcite{krivitsky2009representing}, \textcite{Krivitsky2008}, \textcite{Salter-Townshend2013661}, \textcite{Ryan201770}, \textcite{Gormley2010385}, \textcite{sewell2017latent}, \textcite{aliverti2019spatial}.
This model integrates the latent space distance models with model-based clustering of the nodes, hence connecting to the literature on stochastic blockmodels \parencite{wang1987stochastic, article_Snijders_Nowicki}. 

In the paper, the latent distance model is given as
\begin{equation}\label{eq:lpm_handcock_1}
\begin{aligned}
	{\rm logodds}(y_{i,j}=1|z_i,z_j,x_{i,j}, \beta)
	= \beta_{0}^{\prime} x_{i,j}- \beta_{1} |z_i - z_j|  .
\end{aligned}
\end{equation}
The prior structure is modified by assuming that the latent positions $z_i\in \mathbb{R}^{d}$ arise from a finite mixture of multivariate normal distributions with $G$ components:
$$ z_i \sim \sum_{g=1}^{G} \lambda_g MND_d(\mu_g,\sigma_g^{2}I_d),    $$
where $\lambda_g$ is the probability that a node belongs to the $g$-th group, and $\sum_{g=1}^{G} \lambda_g =1$.
The proposed model captures network features such as transitivity and homophily, but is also capable of representing clustering in a more explicit and natural way.

This work introduces a model-based criterion to select the best number of clusters $G$ and the number of latent dimensions $d$ jointly.
The criterion is inspired by the BIC (Bayesian Information Criterion, \cite{fraley1998many}) and it combines two principled approximations: one determined by $d$, the dimension of the latent space and one determined by $G$, the number of latent clusters or mixture components. 
While this addresses a gap in the literature by providing a principled approach for model choice, it also emphasizes that the method itself (in particular, the BIC) may not be ideal for the selection of the number of dimensions $d$ since the underlying asymptotic approximation result has not been shown to hold in this case.

\subsection{Latent distance models with node-specific  random effects}
\textcite{krivitsky2009representing} combined the approaches of \textcite{Handcock2007301}, which involves model based clustering of the latent space positions, and \textcite{hoff2005bilinear}, which uses actor-specific random effects, into a new model called the Latent Position Cluster Random Effect model (LPCMRE). The proposed model is
\begin{equation}\label{eq:lpm_random_effects_1}
\begin{aligned}
	\eta_{i,j} = {\rm logodds}(y_{i,j}=1|z_i,z_j,x_{i,j},\beta) = \sum_{k=1}^{p}\beta_{k} x_{k,i,j}- |z_i - z_j|+ \delta_{i} + \gamma_{j},
\end{aligned}
\end{equation}
where, in addition to the specification of \textcite{Handcock2007301}, 
$\delta_{i}, \gamma_{j}$ are introduced as actor-specific sender and receiver effects. 
The purpose of these parameters is to represent different levels of sociality of the nodes, so that a wider variety of degree distributions can be represented.

\section{Inference for the LPM}
\subsection{Estimation} 
\textcite{Hoff20021090} considered two main approaches to inference for the LPM. The first is based on maximum likelihood maximization and it consists of a two-step procedure. In the first step, we compute the maximum likelihood estimator of the pairwise distances between all nodes. The log-likelihood is a convex function of the distances between the actors, so it can be easily maximized using numerical procedures to obtain an estimator. In the second step, the optimal distances are used to derive the actual positions of the nodes in the latent space. The authors proposed performing this step using multidimensional scaling. The two-step procedure does not guarantee that the global maximum of the likelihood is obtained, but empirically it has given good results with low computational demands. This is appealing as the set of inferred positions may be used as a starting point for potentially better, but more computationally intensive procedures.

The second approach proposed by \textcite{Hoff20021090} is a Bayesian approach based on Markov chain Monte Carlo (MCMC). 
Prior distributions are specified for the model parameters, and then MCMC is used to obtain approximate samples of the parameters from the posterior distribution.
For the distance model, the prior distributions are typically specified as:
\begin{equation*}\label{eq:prior_1}
\begin{split}
    & z_i \overset{\text{iid}}{\sim} MVN(0, \Sigma) ,  \\
    & \theta_k \overset{\text{iid}}{\sim} N(0, \sigma^2) ,
\end{split}
\end{equation*}
for some covariance matrix $\Sigma$ and variance parameter $\sigma^2$.

The prior distribution is combined with the likelihood function, namely
\begin{equation}\label{log_likelihood}
\mathbb{P}(Y|Z,\theta) = \prod_{i \neq j} \mathbb{P}(y_{i,j}=1|Z,\theta)^{y_{i,j}} \mathbb{P}(y_{i,j}=0|Z,\theta)^{1-y_{i,j}} = \prod_{i \neq j} \frac{\exp{(y_{i,j}\eta_{i,j})}}{1+\exp{(\eta_{i,j})}}  ,
\end{equation}
to obtain the posterior distribution of the model:
$$
\pi(Z,\theta|Y) \propto  \mathcal{L}_Y(Z,\theta)  \pi(\theta) \pi(Z).
$$

The MCMC approach is a standard Metropolis-within-Gibbs algorithm with random walk proposals, where each parameter of the model is sampled in turn from its full conditional distribution. 
The full conditional distributions for the parameters are not in standard form and are given by:
\begin{equation}\label{eq:full_conditional_1}
\pi(z_{i}|Z_{-i},\theta,Y) \propto \pi(z_{i}) \prod_{i \neq j} \mathbb{P}(y_{i,j}=1|Z,\theta)^{y_{i,j}} \mathbb{P}(y_{i,j}=0|Z,\theta)^{1-y_{i,j}},  
\end{equation}
\begin{equation}\label{eq:full_conditional_2}
\pi(\theta_{k}|\theta_{-k},Z,Y) \propto \pi(\theta_{k})  \mathcal{L}_Y(Z,\theta),  
\end{equation}
where the negative subcripts indicate the collection of parameters with the exception of the one negated. 

One challenge with LPMs is the high computational complexity, as highlighted by Eq.~\ref{eq:full_conditional_1} and Eq.~\ref{eq:full_conditional_2}. These updates make the number of calculations that are required for the procedure grow with the square of the number of nodes. 
This quadratic cost is not scalable and it can limit the applicability of the original model.

The MCMC sampling proceeds as follows:
\begin{enumerate}
    \item Choose an initial guess for all the model parameters $Z$ and $\theta$.
    \item For every node $i = 1,\dots,N$:
    \begin{enumerate}
        \item Sample a vector $z^{*}_{i}$ from a multivariate Gaussian proposal $q(z^{*}_{i}\rightarrow z_i)$ which is centered in the current position of this node.
        \item Calculate the ratio between the full-conditionals $r_Z=\frac{\pi(z^*_i|Z_{-i}, \boldsymbol{\theta})}{\pi(z_i|Z_{-i}, \boldsymbol{\theta})}$.
        \item The new proposed values are accepted with probability $\min(1,r_Z)$, otherwise the current values are retained in the sample.
    \end{enumerate} 
    \item For every node $k = 1,\dots,K$:
    \begin{enumerate}
        \item Sample a new parameter $\theta^{*}_{k}$ from a Gaussian proposal $q(\theta^{*}_{k}\rightarrow \theta_k)$ which is centered in the current value of this parameter.
        \item Calculate the ratio between the full-conditionals 
        $r_\theta=\frac{\pi(\theta^*_k|\boldsymbol{\theta}_{-k}, Z)}{\pi(\theta_k|\boldsymbol{\theta}_{-k}, Z)}$.
        \item The new proposed value is accepted with probability $\min(1,r_\theta)$, otherwise the current value is retained in the sample.
    \end{enumerate} 
\end{enumerate}
This algorithm generates a Markov chain whose stationary distribution is the posterior distribution sought.

\subsection{Interpretation of the posterior samples}
The likelihood defined in \textcite{Hoff20021090} depends on the latent positions only through the pairwise distances between the nodes. Hence, the model parameters are non-identifiable with respect to any distance-preserving transformations of the latent positions. These transformations include rotations, reflections and translations of the latent positions. 
In principle, if we focus only on one single configuration of the parameters drawn from the posterior distribution, the non-identifiability issues may be negligible, because a translation, reflection, or rotation of the points would not affect our interpretations of the latent space.

However, in a Bayesian setting, we obtain a posterior sample $\{Z^{(1)}, Z^{(2)}, \dots\}$, and we are interested in summarising that sample to obtain estimators such as the posterior mean.
In this case, applying a transformation (e.g. a translation) to any of the configurations of the sample would directly affect the value of the posterior summaries.
Crucially, we have no way to determine whether the latent space was rotated, translated or reflected during the sampling procedure.
For this reason, the posterior sample is not identifiable, and we cannot draw meaningful summaries from it.

As a solution to this problem, \textcite{Hoff20021090} considered Procrustes matching. This approach is based on a comparison of one configuration of points with another via their respective coordinate matrices (we denote these matrices by $A$ and $B$ in this section). In Procrustes matching, a rigid transformation is applied to $A$ to make it as close as possible to the reference $B$. The matrix $A$ is rotated, reflected and translated to achieve the
best match to the reference matrix $B$, where the best match is defined as the one that minimises the sum of squared distances between corresponding points.
This is given by:
$$ R^2 =\sum_{i=1}^{n}\sum_{k=1}^{d} (b_{i,k}-a_{i,k})^2. $$

We can set up an optimization problem by specifying the following transformation of the matrix $A$:
$$ \textbf{a}_i' = \textbf{O}^\top \textbf{a}_i + \textbf{t} ,   $$
where $\textbf{a}_i$ is a vector of coordinates of point $i$ in the matrix $A$, and $\textbf{O}$ is an orthogonal matrix that induces a rotation and/or reflection and $\textbf{t}$ is a translation vector. 
Then, the optimal values of $\textbf{O},$ and  $\textbf{t}$ can be calculated by minimizing the sum of squared distances between points, namely
$$ R^2 =\sum_{i=1}^{n} (\textbf{b}_{i}-\textbf{O}^\top \textbf{a}_i + \textbf{t})^\top  (\textbf{b}_{i}-\textbf{O}^\top \textbf{a}_i + \textbf{t}). $$

The optimal values obtained are then applied to the original $A$ configuration leading to the minimum $R^{2}$ value (called the Procrustes sum of squares). The computation for the projection model is slightly different from that for the distance model.
Since the projection model is not invariant under translation, the optimization problem involving transformation of matrix $A$ reduces to:
$$ \textbf{a}_i' = \textbf{O}^\top \textbf{a}_i.  $$
Thus the modified sum of squared distances is 
$$ R^2 =\sum_{i=1}^{n} (\textbf{b}_{i}-\textbf{O}^\top \textbf{a}_i)^\top  (\textbf{b}_{i}-\textbf{O}^\top \textbf{a}_i).  $$

The problem of identifiability has been further studied by \textcite{Shortreed200624} to extend the work of \textcite{Hoff20021090}. 
They considered three point estimators for the positions of the nodes, namely, the maximum likelihood estimator, the posterior mode estimator, and the posterior mean estimator (defined after Procrustes matching). 
They argued that these three estimators may provide inaccurate point estimates of the nodes, since they rely exclusively on the estimated distances to derive the positions. 
They introduce an original fourth procedure which aims at minimizing the Kullback-Leibler divergence between the distribution implied by the fitted positions, and the one implied by the parameters minimizing the expected posterior loss.
This fourth approach is most commonly used \parencite{Handcock2007301, krivitsky2009representing} and is implemented in the \texttt{latentnet} \texttt{R} package.

\subsection{Other inferential approaches}
The computational complexity of the log-likelihood function in Eq.~\ref{log_likelihood} is on the order of the square of the number of nodes. 
This is a problem when inferring LPMs, because likelihood-based procedures require a quadratic computing cost, which does not scale well with the size of the data.
The full Bayesian approach based on MCMC is particularly slow and becomes impractical for networks of more than a few hundred nodes.
On the other hand, the MLE approach tends to be faster, yet it still requires a quadratic cost and thus it too does not scale well. 

A number of papers have specifically addressed this research impasse, by proposing various strategies that can speed up the inferential procedures and thus make the algorithms scale better with large datasets.
The variational Bayesian inference procedure proposed by \textcite{Salter-Townshend2013661} proposes to replace the posterior distribution of the model parameters in the Latent position cluster model with a variational posterior. 
This leads to an approximation that has been widely used in a variety of different settings with latent variables \parencite{jordan1999introduction, blei2017variational}.
The advantage of this approximation is that it allows one to characterize the posterior distribution of interest without having to resort to sampling methods like MCMC, which can be much more computationally intensive.
On the other hand, the approximation error induced is difficult to quantify, and it has been shown that this error can be relevant when assessing the uncertainty around the point estimates provided. 

Another approach is considered by \textcite{Raftery2012901}, who
introduce a likelihood approximation using case-control sampling. 
This takes advantage of the fact that the positions of the nodes are primarily regulated by the edges between nodes, rather than by the missing edges. This is especially true in sparse networks, where the few edges that are present give us information on which nodes should be located close to each other. By contrast, the missing edges tend to not provide as much information, and so can be thinned out using a case-control sampling strategy. 
As a consequence, only a fraction of the missing edges are considered in the likelihood calculation, and their contributions are reweighted to make them represent all the missing edges.
This approximation yields an unbiased estimate of the likelihood function, while scaling up the procedure to networks of several thousand nodes. 

\textcite{Ryan201770} extend the distance LPCM by including conjugate priors in the model specification. The key advantage of this approach is that a number of model parameters can be integrated out from the posterior distribution thanks to the conjugacy property. 
This leads to a marginal posterior which can be characterized through a MCMC sampler. 
The advantage of this approach is that it can speed up the sampling procedure since many of the model parameters have been integrated out from the posterior. 
The sampler can explore different models with a different number of latent clusters in one run. This is a key advantage because it allows one to perform model choice on the number of groups without having to fit each of the models individually.

These methodological ideas  were introduced for binary graphs, but they have been extended and adopted in a variety of different settings with dynamic and multi-view networks.
More recently, some other methodological ideas have been considered \parencite{aliverti2021stratified, rastelli2018computationally, spencer2020faster, liu2021variational} signalling that this research problem remains one of active interest.

\section{An illustrative example for the distance LPM}\label{sec:example}
We now describe a motivating example on an original coauthorship network. 
The data (collected from Scopus on January 17th, 2022) consists of all the bibliographic entries which cite \textcite{Hoff20021090}.
A total of $929$ articles are part of this dataset, and 1,758 unique authors are involved. 
From these, we constructed a coauthorship network for all the authors of those papers, where two authors are linked if they coauthored at least one paper.
We removed all the authors that published only one paper in total, to focus on the most active authors and to reduce the computational burden. 
This led to a final undirected binary network of $279$ nodes. 

\begin{figure}
    \centering
    \includegraphics[width = 0.7\textwidth]{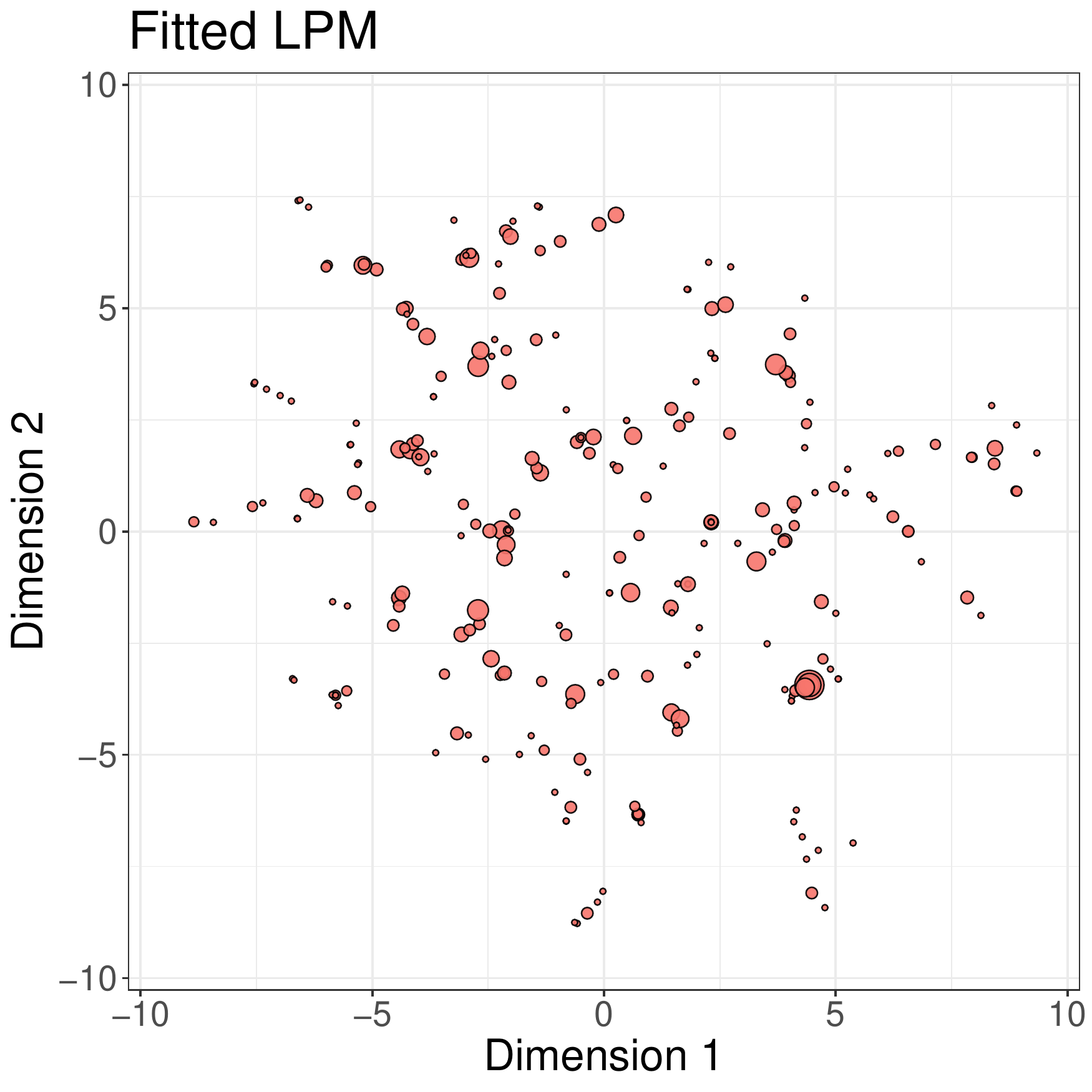}
    \caption{Distance LPM fitted on the coauthorship data. Each node represents a different author. The size of a node represent the number of publications in which the author participated.}
    \label{fig:coauthorships_1}
\end{figure}

We then fit a distance LPM on these data using the \texttt{R} package \texttt{latentnet}.
We compared a number of models using the BIC criterion, as implemented in \texttt{latentnet}. 
The resulting latent space is shown in Figure~\ref{fig:coauthorships_1}, and
a more detailed representation of the results through an interactive plot is publicly available on GitHub. 

\footnote{URL for GitHub: https://github.com/riccardorastelli/CoauthorshipsLPM}

\begin{figure}
    \centering
    \includegraphics[width = 0.495\textwidth]{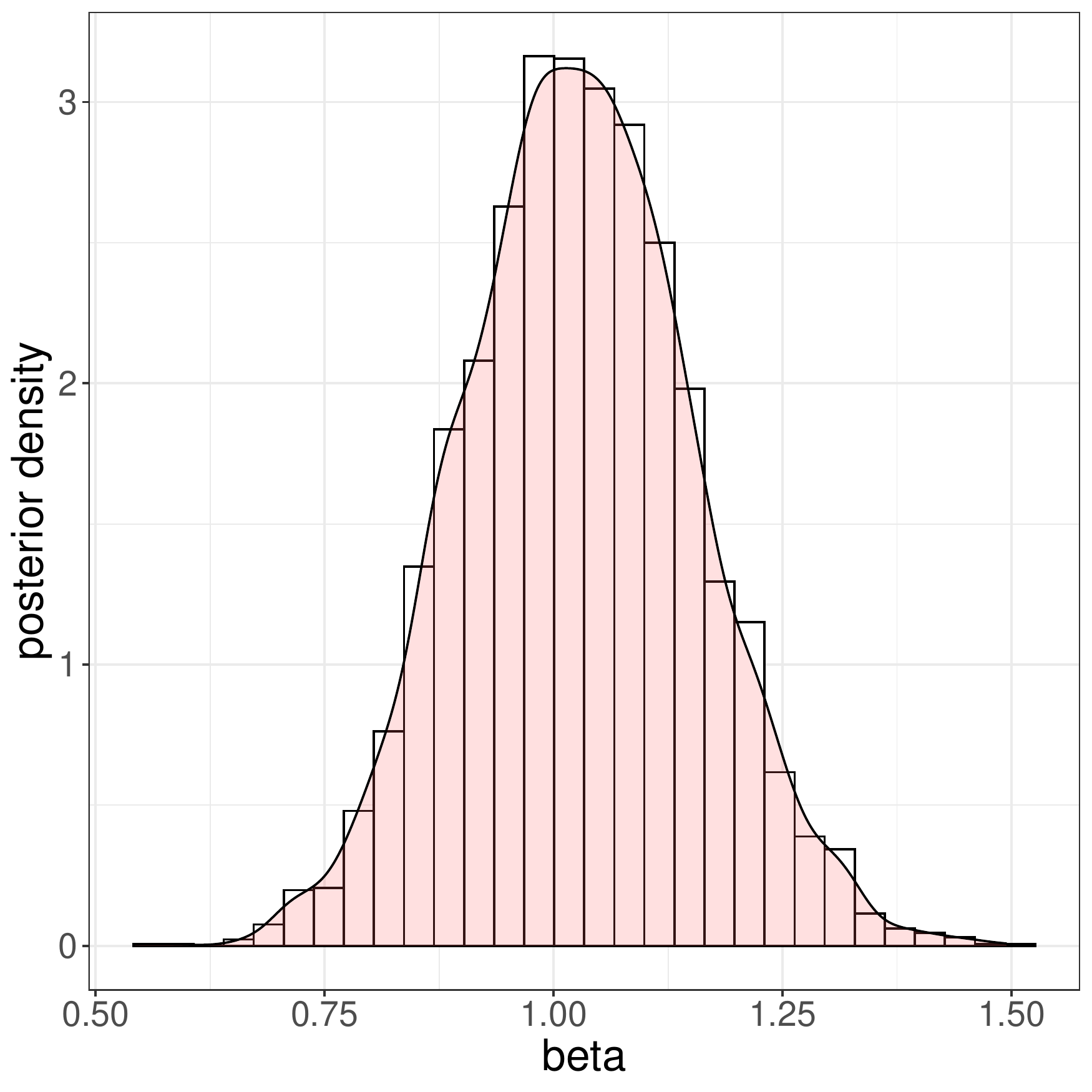}
    \includegraphics[width = 0.495\textwidth]{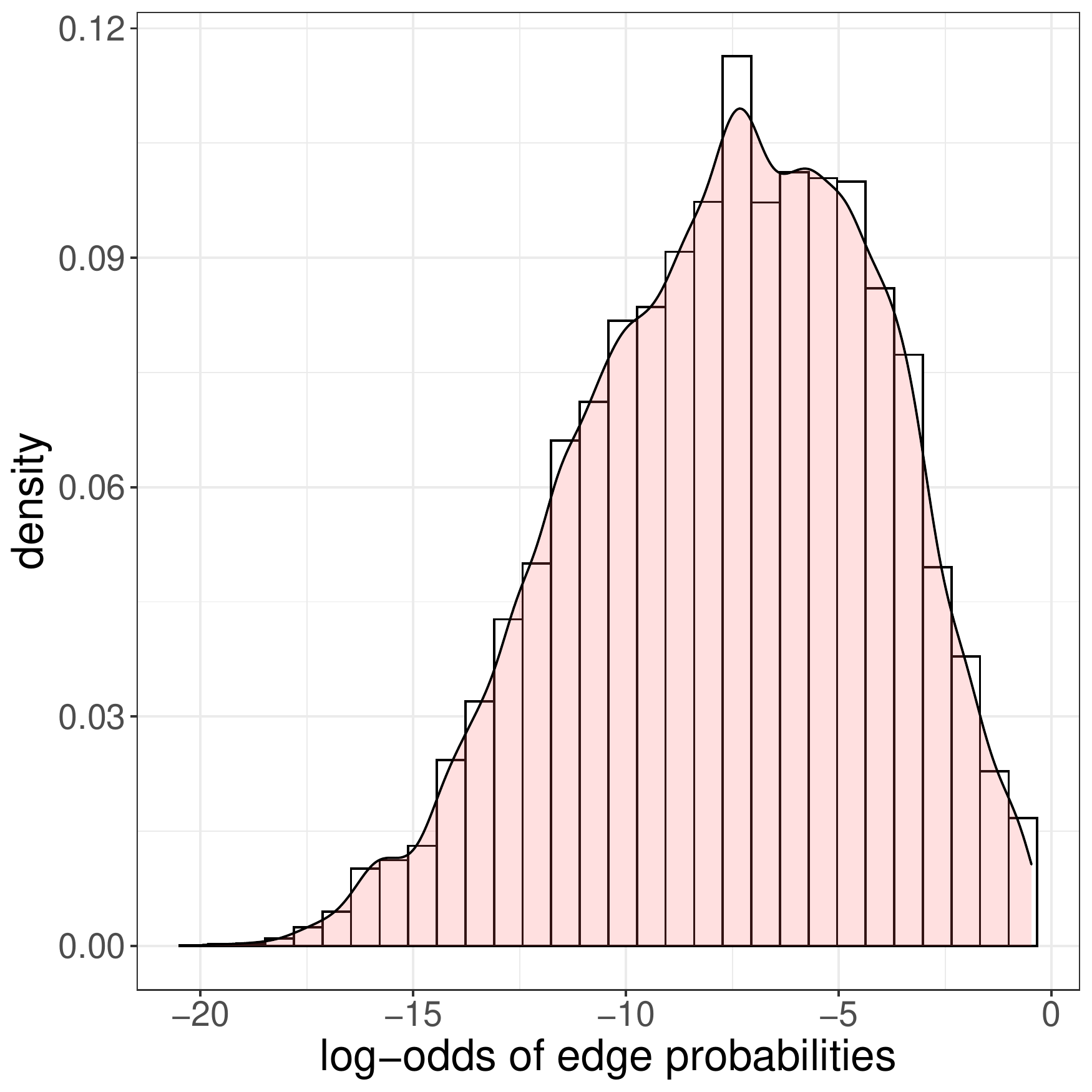}
    \caption{The left panel shows the approximate posterior sample for the intercept parameter, called $\beta$. The right panel shows instead the $\eta$ parameters representing the log-odds for the edges appearing. The distribution represents these values across all dyads.}
    \label{fig:coauthorships_etabeta}
\end{figure}

The results highlight a strong presence of communities, which reflect the presence of various research groups.
Some of the authors tend to have overlapping positions, suggesting strong similarity of research collaborators. Others are positioned in between communities, highlighting that they may have connections with more than one research group or with a more diverse set of collaborators.
In Figure~\ref{fig:coauthorships_etabeta} we show the posterior sample for the intercept parameter, and the distribution of the log-odds for all edges.

These plots show that the latent space plays an important role in determining the presence of edges. The left panel shows that the intercept parameter concentrates around the value $1$, signalling a strong effect of the latent space. This is confirmed on the right panel of the same figure, where we can see that the log-odds exhibit large variability, implying edge probabilities that range from $0.00$ to $0.38$. This means that the model is able to represent the heterogeneity in the data by leveraging the geometric framework induced by the LPM.

\begin{figure}
    \centering
    \includegraphics[width = 0.7\textwidth]{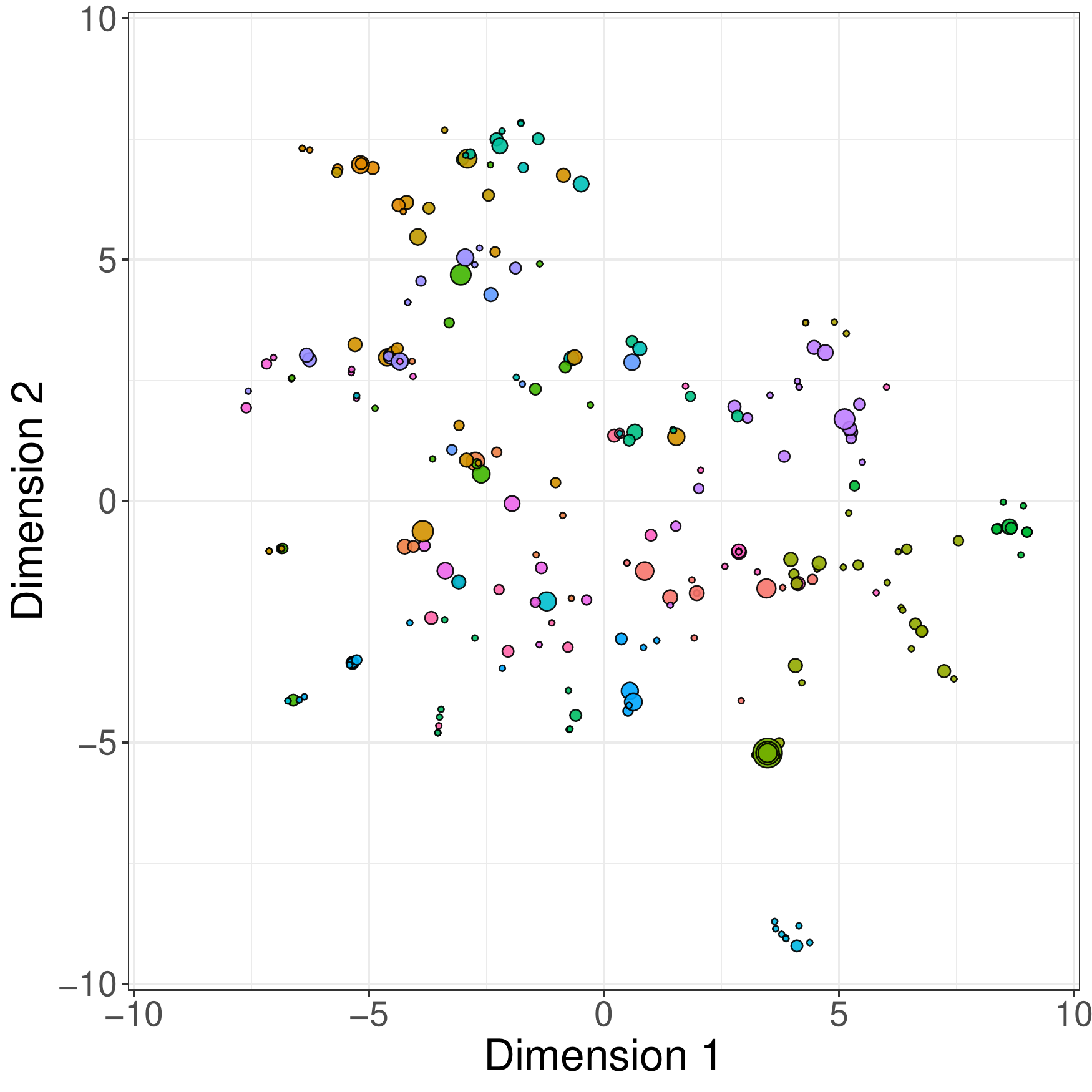}
    \caption{Distance LPCM fitted on the coauthorship data. The size of the nodes represent the total number of publications to which the authors participated. The colors of the nodes represent their partitioning.}
    \label{fig:coauthorships_2}
\end{figure}

In the fitted LPM of Figure~\ref{fig:coauthorships_1} we can explore the presence of communities by studying how ``clustered'' the latent space is.
However, the LPCM allows one to include this information directly in the modelling, and thus obtain a partitioning of the nodes into groups.
For this reason, we also fitted a LPCM with up to $30$ groups, and selected the best model using the BIC criterion advocated by \textcite{Handcock2007301}.
The results ($27$ groups) are shown in Figure~\ref{fig:coauthorships_2}.
In this case we notice strong agreement between the partitioning of the nodes and the clustering of points in the latent space. 
The research groups are fairly well captured and they convey an interesting model-based view of the data.

\begin{figure}
    \centering
    \includegraphics[width = 0.7\textwidth]{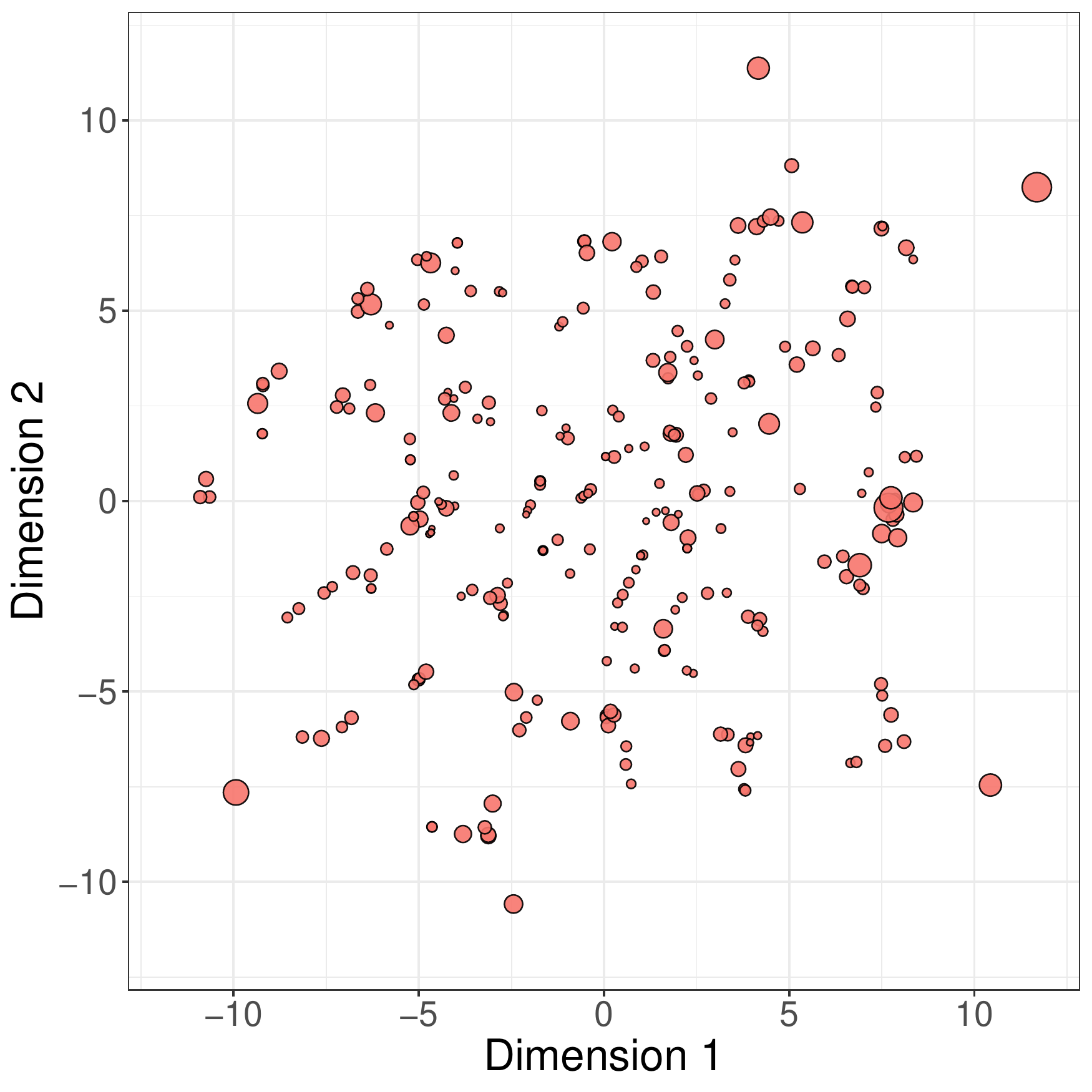}
    \caption{Distance LPCMRE fitted on the coauthorship data. The size of the nodes represent their "sociality" random effect parameter.}
    \label{fig:coauthorships_3}
\end{figure}

In addition, we also provide a third example using the model of \textcite{krivitsky2009representing}.
Also in this case we fit the model for different number of groups and we select the best fit using BIC.
The results for the best model (shown in Figure~\ref{fig:coauthorships_3}) are obtained when only one group is chosen.
This is reasonable since the random effects used in the model can explain a substantial part of the variability in the data, at the expense of the latent space. Nonetheless, the visualization that the model can offer still includes all the information regarding the model parameters since the sociality of nodes is shown through their size.

In this case, we observe that the sociality is similar for most of the nodes that are positioned in the center of the latent space. 
These nodes are well characterized by their latent positions, and the model does not require any random effect to ``correct'' their connectivity patterns. 
Some nodes with a relatively central position also benefit from a high sociality parameter.
This represents the fact that some authors will collaborate with diverse research groups and may require the creation of long-distance edges.
The sociality parameter addresses this situation, helping to bridge that gap and so by helping to create connections regardless of the pairwise distance.
On the other hand, some of the largest sociality values are observed for nodes in the fringes of the latent space.
These authors do not have many connections; for this reason it is difficult to position them centrally or close to other nodes.
As a consequence they are pushed to the edge of the latent space. However, since they do have connections to more central nodes, a large sociality parameter can provide a compensation for their loose positioning thus well justifying their observed connections.

\section{State of the art}\label{sec:literature_review}
The work of \textcite{Hoff20021090} has led to a strand of literature that focuses on LPMs and their extensions. In this section, we review some methodological contributions that extend the work of \textcite{Hoff20021090} in different directions.
We organize our exposition by categorizing the papers into the type of network data they refer to. However many of the papers are generalizable to multiple contexts and provide novelty in a number of different ways.

\subsection{LPMs for multiview networks}
Multiview or multiplex networks are observed when there are multiple relationships among the same set of nodes, hence describing a multi-relational network.
An example of this framework may be a social network where we observe various types of ties between actors, such as professional ties, family ties or friendship ties.
Also, temporal networks are closely connected to multi-view networks as they may be seen as a special case where a time dependency is created to connect contiguous network views.

LPMs have been extended to the multi-view framework in a number of papers.
\textcite{Gollini2016246} introduced the latent space joint model which displays the different relations on the same nodes to be encapsulated in the same latent space. 
This also introduces a scalable variational approach to speed up the estimation of the latent space model.
Another novelty is that they use the squared Euclidean distance instead of the absolute distance in their model.

Instead, \textcite{Durante20171547} develop an extension of the latent projection model of \textcite{Hoff20021090} using a novel non-parametric Bayesian approach to model the latent space and the clustering of the nodes. They propose an application to a population of brain connectivity networks, whereby nodes refer to anatomical brain regions and edges refer to the structural interconnections between them. They consider data collected from a number of different samples/individuals, which is where the multi-view aspect of their analysis comes from.
The application of their model can be seen in a subsequent paper \parencite{Durante201829} in which the model is used to characterize and test for differences in brain connectivity networks. 
A second paper in a similar direction is \textcite{aliverti2019spatial}.

\textcite{Salter-Townshend20171217} propose a new model for multiview networks which focuses on the interrelation of dyads across different network views. 
They use a multivariate Bernoulli likelihood whose parameters are determined by multiple latent spaces and by the distances between nodes in different views.
The framework permits a model-based assessment of the correlation between the multiple edges, providing a measure of association between the various network views.
The price to pay for this additional flexibility however, is increased computational requirements and a more challenging model-choice task. 

\textcite{angelo2019900} and \textcite{D_Angelo2020324}
introduce a Euclidean distance LPM for multi-view networks to model the voting patterns in the Eurovision contest.
The authors specify a common latent space to represent the positions of the nodes, but they let the other parameters be different across the various network views. 
This is a key aspect of the model since it directly allows one to quantify the effect that the latent space can have on the generation of the data.
Similarly to Eq.~\ref{eq:lpm_handcock_1}, a coefficient for the latent distances approaching zero would mean that the latent distances do not represent a strong effect, and thus the data can be well represented through a homogeneous model.
The authors use this aspect of the LPM to determine the effect of the latent space as a way to measure the voting bias of the countries.

\textcite{Sewell2019160} introduce a new method to analyze cognitive social structures based on the LPM. These data refer to the perception that each node has on the network, so it is related to multi-view networks in that the observed data corresponds to a collection of networks. 
He derives an LPM approach which combines the advantages of the latent space visualization, with additional model parameters which can measure the nodes' biases in perceiving the network.

\subsection{LPMs for dynamic networks}
Dynamic or temporal networks refer to networks that evolve over time.
Most typically, dynamic networks describe the interactions between the same set of nodes, and are observed through network snapshots, i.e. at some specific points in time. For this reason, they can be formally represented as a collection of graphs $Y^{(1)},\dots,Y^{(K)}$. 
The LPM of \textcite{Hoff20021090} has been extended to dynamic networks in a variety of ways.
\textcite{Sarkar20051145} extend it to include time-dynamics in the model parameters. They propose an application to a dataset of friendship relationships and study how these change over time. 
The proposed model embeds the nodes in the latent space in such a way that the positions of nodes are regularized at each time point to ensure that their positions do not change by much. 

\textcite{Sewell20151646} introduced a new model to extend that of \textcite{Sarkar20051145}. 
In their framework, the longitudinal network data is modeled via temporal trajectories in a latent Euclidean space. 
One key novel aspect of this work is that the formulated model introduces a social reach parameter which resembles the nodal random effects of \textcite{krivitsky2009representing}. These parameters aim at representing different levels of activity of the nodes and their sociability, regardless of the latent position. 

\textcite{Friel20166629}. 
define a model for bipartite dynamic networks, and apply their method to a network of companies and their directors in Ireland. 
The model aims at providing a model-based assessment of how appointments of directors to multiple boards may have been associated with financial instability during the 2008 crisis.
The authors embed the nodes of bipartite networks into a single latent space, similarly to \textcite{Gormley200790}.
They assume that the nodes can move along trajectories, hence capturing the time dependencies in the data. 
In addition, they note that the data present a strong persistence of edges and non-edges. 
This refers to dyads that tend not to change their status over time. In order to address this aspect of the data, which is common to many dynamic network datasets,  the authors also introduce a regime-switching framework which selects a different intercept parameter based on the previous network realization. 
The model captures the heterogeneity shown by the data very well through latent positions and persistence features by modeling the structure of the intercept parameters. 

\textcite{Durante20162203} further extend the LPM by allowing the positions of nodes to evolve continuously over time through a nested Gaussian process, which results in time-varying smoothness. The authors incorporate locally adaptive dynamics to study face-to-face dynamic contact network, highlighting the flexibility of their approach in capturing data observed at unequally spaced intervals and structural changes in the nodes trajectories.

In a similar direction, \textcite{rastelli2021continuous} introduce a fully continuous latent position model, which extends the application of LPMs to time-stamped interaction data. They consider and infer the continuous trajectories of the nodes in a two-dimensional latent space, using only their frequency of pairwise connections.

\subsection{LPMs for weighted networks}
Weighted networks are characterized by edges that carry additional numerical information. 
Typically, the edges of these networks carry a real value, for example representing the intensity or the length of the interaction. 
Although binary networks are easier to study, weighted networks arise often in practice.
\textcite{hoff2005bilinear, hoff2021additive} proposed a general framework
that includes various types of latent variable models as special cases, including the distance LPM and the projection LPM.
The network variables are considered as the response variables of a regression model, and are characterized by a combination of nodal effects, dyadic effects and other types of network-derived information. 
\textcite{westveld2011mixed} extended this framework to model a network of international trade and conflict.

\textcite{Sewell2016105} introduce a dynamic weighted network framework. 
They define the temporal dependencies through a Markovian property, and consider two models for the weights: a Poisson model for count data and a Tobit model for nonnegative continuous data. 
They also consider nodal random effects similarly to \textcite{Sewell20151646}, to define a possibly different social reach for each node.

\subsection{LPMs for rank data}
Ranked network data can be used to represent rankings of items or entities by the nodes. 
A bipartite network can be used to represent how a set of nodes rank various items, or the other nodes themselves.
\textcite{Gormley200790} combined the LPM of \textcite{Hoff20021090} with the Plackett-Luce model \parencite{plackett1975analysis} to create a new type of LPM which can be used to study bipartite rank networks, and applied their method to the voters' preferences in the 2002 Irish elections.

\textcite{Sewell2015611} further extended the approach of \textcite{Gormley200790} to account for temporal changes in the networks as well as a social reach effect on the nodes. 
Their LPM is dynamic in that the nodes move along trajectories, and their positions are used to characterize the various network snapshots at different times. This permits an analysis of the evolution of the latent space, and thanks to the node-specific parameters, it allows for an assessment of the popularity and stability of individual actors. In both papers, inference is carried out using Bayesian framework and Procrustes matching. As concerns scalability, the methods proposed are particularly demanding since they require an even higher complexity than the original LPM. This points to an important research question for future work in this direction.

\subsection{Theory of LPMs}
The papers we consider in this section are general in that they are not restricted to a particular network type.
We first highlight a number of theoretical papers that have studied the LPM framework or some closely connected variants.
\textcite{Rastelli2016407} provide a detailed analysis of the LPM framework by considering some theoretical and empirical aspects that arise from the model structure. They address how the LPM can capture some relevant network features of interest such as clustering, heavy-tailed degree distributions, small world behavior, and assortative mixing. 
They also propose a variant of Eq.~\ref{eq:lpm_hoff_2} in which the logistic expression is replaced by a Gaussian kernel. This choice permits a detailed analysis of the theoretical properties of the model, creating a connection to the physics literature on the topic \parencite{newman2018networks}.

\textcite{Smith2019428} focus on the role played by the geometry of the latent space in a LPM. They create a general framework which includes a variety of latent variable models, and they study how switching to a hyperbolic geometry can lead to more flexible network models which can better represent the observed data. In particular, they show that a negative curvature in the space permits more complexity without the need to change any other aspect of the models, hence providing a more parsimonious modeling choice. 

\textcite{caron2017sparse} consider a general latent variable framework, and argue that commonly used frameworks such as the LPM may not be adequate for large networks since they cannot capture sparsity in asymptotic settings. They then propose a new framework based on exchangeable random measures which is capable of maintaining ideal features like network sparsity along with exchangeability properties.

\subsection{Extensions of the projection LPM}
The distance LPM and the projection LPM introduced by \textcite{Hoff20021090} generated two quite separate sets of literature. 
On the one hand, the distance LPM has been widely applied thanks to its clear representations and easy interpretability.
On the other hand, the projection model has been intensely studied due to its tractability and also its similarity to other types of latent factor models.
In the projection LPM literature, the work of \textcite{hoff2005bilinear} has initiated a research avenue \parencite{hoff2007modeling, hoff2011hierarchical, hoff2021additive} which has evolved into a flexible and widely applicable framework for network analysis.

\textcite{nickel2008random} and \textcite{young2007random}
introduce a general latent variable multiplicative structure which is closely connected to the projection model, and call this the Random Dot Product graph model.
In its original formulation, the model does not include the logit link for the multiplicative effects.
However, this becomes an advantage because it makes the model quite tractable, and so a number of network properties can be derived analytically.
A number of recent papers extend the random dot product graph and use this framework in network applications \parencite{athreya2021estimation, sanna2021link, ng2019generalized, xie2021efficient, Zhang202011225}.

\subsection{Extensions of the LPCM}
The distance LPCM of \textcite{Handcock2007301} has been an especially influential since it inherits the advantages of the basic LPM of \textcite{Hoff20021090} while gaining more flexibility and interpretability through the clustering structure.
This has led to a rich literature about this model and its applications.

\textcite{fosdick2019multiresolution} introduce a new latent variable network model inspired by the LPCM. They argue that the complexity of networks may be better captured when different statistical models are combined at different level of resolution. 
This means that the models may be either stacked into different hierarchical layers, or they may model different aspects of the data.
They illustrate their approach by proposing a new model, the latent space stochastic blockmodel, which combines the stochastic blockmodel of \textcite{wang1987stochastic} with the LPM of \textcite{Hoff20021090}.
Nodes are clustered into groups following a stochastic blockmodel structure, while the probabilities of connections for nodes belonging to the same group are determined by a group-specific LPM.
This effectively creates different resolutions in the model where the inner level, represented by the LPM, can break down and capture the within-group heterogeneous connectivity patterns.
The result is a framework that can scale better than that of \textcite{Handcock2007301} while maintaining the flexibility and interpretability of the LPMs.

Another approach which combines LPMs and clustering is that of \textcite{Sewell2020390}.
He considers a clustering framework on the edges rather than on the nodes. 
This reflects the idea that edges are determined within a particular context they refer to, for example because the nodes interacted within that context.
He then introduces a latent space to formalize the presence of these contexts that drive the formation of the edges.
As a result, the nodes involved in a particular edge are chosen based on their relative positions with respect to the context of the edge.
The methodology can provide latent space representations (as well as clustering) of edges and nodes, hence creating a strong connection with the LPCM.
However, differently from the LPCM, the computational complexity of the approach proposed scales with the number of edges in the graph, so it may be particularly suitable for sparse networks.

A dynamic extension of the LPCM is introduced by \textcite{sewell2017latent}. In this work, the authors consider a dynamic LPM where both the latent positions of the nodes, and their cluster allocations, can change over time. The the position of a node is influenced by its previous position and its previous cluster membership. The clustering variables are assumed to have a Markovian structure and depend on their previous value. The authors consider an estimation strategy based on variational inference (for large networks), and another strategy based on MCMC (for smaller networks due to higher computational costs). Also, they perform model choice using BIC to select the number of clusters, and DIC to select the number of latent dimensions.

\textcite{Gormley2010385} extend the framework of \textcite{Handcock2007301} to include node covariates. While the LPCM can accommodate edge covariates directly into the edge probabilities through the logit link, the inclusion of a node's information is not as straightforward. \textcite{Gormley2010385} address this gap by proposing a mixture of experts model as a prior for the latent positions, whereby the nodes' covariates can be used to characterize the group allocation probabilities associated to the nodes.
This defines a framework in which covariates can be introduced both via the likelihood, and also into the structure of the model, providing more flexibility and possibly a different interpretation of the data.

\section{Software}
An essential aspect of latent position modeling is the implementation of the methods, which include MCMC, variational methods and other procedures. 
MCMC estimation of a range of latent position network models is carried
out by the \texttt{R} package \texttt{latentnet} \parencite{Krivitsky2008,krivitsky2020package}.
This implements the methodologies outlined in \textcite{Hoff20021090, Shortreed200624, Handcock2007301, krivitsky2009representing}. 
This package has been used to produce the examples in this paper.

\section{Open research questions}
LPMs have had great success in the last two decades thanks to their ability to capture some common network properties, such as transitivity, homophily and clustering. Research on these models is active, and several methodological questions remain open.

One key issue concerns computational efficiency and scalability. While this is a relevant topic throughout network science, it is especially important in the context of LPMs.
As mentioned earlier, the original LPM has a quadratic cost which arises from the likelihood calculation. 
The model requires information from each dyad of the network, and this dyadic information is possibly always different, generating the quadratic cost.
While a number of ideas have been proposed, the current approaches can only scale to thousands of nodes, whereas observed datasets can be much larger. 
This gap highlights the need to define models and methods that can take computational efficiency directly into account, while providing a solid structure for statistical analyses.

Another aspect of the research on LPMs refers to its flexibility in capturing features of interest.
A closely related latent variable model, the stochastic blockmodel, can capture both assortative and disassortative behaviors of the nodes. 
The original LPM falls short in this regard, since the model can capture assortative mixing but it cannot capture disassortative mixing.
While social networks tend to often show assortative mixing, other types of networks such as biological networks can exhibit strong disassortative patterns \parencite{newman2018networks}.
For this reason, variations of the LPM have been recently considered to make the model more flexibile in this regard, for example by combining it with the stochastic blockmodel \parencite{fosdick2019multiresolution, ng2021modeling, rastelli2018sparse}. 

Another key open research question relates to model choice, and, in particular, to the selection of the dimensionality of the latent space. 
Most commonly, LPMs are fitted in two latent dimensions, due to easier interpretations and clearer visualizations.
However, a number of papers have highlighted how the choice of this parameter, the number of latent dimensions, can be set up as one of model choice.
A BIC criterion was introduced by \textcite{Handcock2007301} to select the number of clusters and the number of latent dimensions jointly. However, the authors note that the asymptotic properties of BIC may not be valid when selecting the number of dimensions.
A Deviance Information Criterion (DIC, \cite{spiegelhalter2002bayesian}) has been used by \textcite{Friel20166629, angelo2019900, sewell2017latent}.
However, more research on the model choice issue is needed.

\section{Acknowledgements}
This publication has emanated from research supported in part by a grant from the Insight Centre for Data Analytics which is supported by Science Foundation Ireland under Grant number $12/RC/2289\_P2$.

\printbibliography

\end{document}